\pdfoutput=1
\documentclass[prl,twocolumn,preprintnumbers]{revtex4}

\makeatletter
\DeclareOldFontCommand{\rm}{\normalfont\rmfamily}{\mathrm}
\DeclareOldFontCommand{\sf}{\normalfont\sffamily}{\mathsf}
\DeclareOldFontCommand{\tt}{\normalfont\ttfamily}{\mathtt}
\DeclareOldFontCommand{\bf}{\normalfont\bfseries}{\mathbf}
\DeclareOldFontCommand{\it}{\normalfont\itshape}{\mathit}
\DeclareOldFontCommand{\sl}{\normalfont\slshape}{\@nomath\sl}
\DeclareOldFontCommand{\sc}{\normalfont\scshape}{\@nomath\sc}
\makeatother

\usepackage[T1]{fontenc}
\usepackage[utf8]{inputenc}

\usepackage[hidelinks]{hyperref}
\usepackage[parfill]{parskip}
\usepackage[separate-uncertainty = true,multi-part-units=single]{siunitx}
\usepackage{microtype}
\usepackage{csquotes}
\usepackage{graphicx}
\usepackage{amsmath}
\usepackage{amssymb}
\usepackage{todonotes}

\usepackage{multirow}

\usepackage{cleveref}

\newcommand{\beq}{\begin{equation}}
\newcommand{\eeq}{\end{equation}}
\newcommand{\bea}{\begin{eqnarray}}
\newcommand{\eea}{\end{eqnarray}}

\usepackage{scalefnt}
\newcommand{\abbrev}{\scalefont{.9}}

\newcommand{\NNLO}{\text{\abbrev NNLO}}
\newcommand{\NLO}{\text{\abbrev NLO}}
\newcommand{\LO}{\text{\abbrev LO}}
\newcommand{\QCD}{\text{\abbrev QCD}}
\newcommand{\PDF}{\text{\abbrev PDF}}
\newcommand{\LHC}{\text{\abbrev LHC}}
\newcommand{\CMS}{\text{\abbrev CMS}}

\newcommand{\ATLAS}{\text{\abbrev ATLAS}}

\newcommand{\DIS}{\text{\abbrev DIS}}

\newcommand{\LHAPDF}{\text{\abbrev LHAPDF}}

\newcommand{\DDIS}{\text{\abbrev DDIS}}

\newcounter{notecount}

\preprint{FERMILAB-PUB-21-456-T, IIT-CAPP-21-01}

\begin{document}

\title{\LARGE Testing parton distribution functions with\\ t-channel single-top-quark production}

\author{John Campbell}
\affiliation{Fermilab, PO Box 500, Batavia, Illinois 60510, USA}
\author{Tobias Neumann}
\affiliation{Department of Physics, Brookhaven National Laboratory, Upton, New York 11973, USA}
\author{Zack Sullivan}
\affiliation{Department of Physics, Illinois Institute of Technology, Chicago, Illinois 60616, USA}

\vspace{1cm}

\begin{abstract}
	The production of single top-quarks in the $t$-channel at hadron colliders imposes strong 
	analytic constraints on 
	parton distribution functions (\PDF{}s) through 
	its 
	double deeply inelastic scattering (\DDIS{}) form. We exploit this to provide novel 
	consistency checks between \LO{}, \NLO{} and \NNLO{} \PDF{} fits and propose to include 
	it as a constraint in future \PDF{} fits. Furthermore, while it is well-known that the 
	$b$-quark \PDF{} is highly sensitive to the $b$-quark mass, we show that the treatment
	of this systematic uncertainty is still incomplete, fragmented or outright missing at the moment.
	Consequently, we conclude that the $b$-quark mass uncertainty is the dominant but so far 
	broadly 
	neglected theory 
	uncertainty for this process.
\end{abstract}

\maketitle

\section{Introduction}
\label{sec:introduction}

Parton distribution functions (\PDF{}s) are an integral part of high-energy collider predictions.
With increasing proficiency in multi-loop predictions for perturbative hard scattering 
cross-sections, \PDF{}s 
are now one of the largest sources of uncertainty in theory predictions. 
Some processes are initiated 
by heavy-quarks with 
masses large enough to be in the perturbative regime. The construction of associated heavy-quark 
\PDF{}s requires an understanding of quark masses, mass-threshold and mass-scheme effects, all of 
which 
contribute to additional systematic \PDF{} uncertainties. These are typically not covered by 
uncertainty prescriptions of individual \PDF{} groups yet \cite{Accardi:2016ndt}, but have become 
relevant with small percent-level perturbative truncation and \PDF{} fit uncertainties. Through 
\PDF{} sum-rules such heavy-quark effects even propagate to other processes.

\begin{figure}
	\centering
	\includegraphics[width=1.5in,clip]{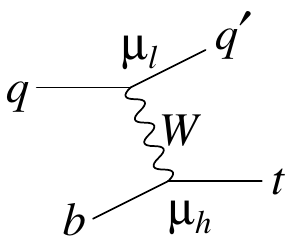}
	\caption{Leading order Feynman diagram for $t$-channel
		single-top-quark production. The light-quark and heavy-quark parts
		of the diagram factorize with independent renormalization- and factorization-scales 
		$\mu_l=\sqrt{Q^2}$ 
		and $\mu_h=\sqrt{Q^2+m_t^2}$, respectively, where
		$Q^2$ is the virtuality of the $W$ boson.}
	\label{fig:tchan}
\end{figure}

One of the most precisely measured heavy-quark initiated processes is $t$-channel single-top-quark
production, see \cref{fig:tchan}. As we elaborate below, $t$-channel 
production plays a pivotal role in the development and understanding
of improved perturbation theory and heavy quark \PDF{}s.
It has been measured inclusively at the Tevatron \cite{CDF:2009itk,D0:2009isq}, 
and also more differentially at the \LHC{} operating at \SI{7}{\TeV} 
\cite{ATLAS:2012byx,ATLAS:2014sxe,CMS:2011oen,CMS:2012xhh}, \SI{8}{\TeV} 
\cite{ATLAS:2017rso,CMS:2014mgj} and \SI{13}{\TeV} \cite{ATLAS:2016qhd,CMS:2016lel,CMS:2018lgn}.
The precision of the cross-section measurement is at the level of 13\% for the Tevatron 
\cite{CDF:2015gsg} and
7\% for the \LHC{} \SI{8}{\TeV} \ATLAS{} and 
\CMS{} combination \cite{ATLAS:2019hhu}.

The analytic connection between $t$-channel single-top
and deeply inelastic scattering (\DIS{}) puts significant constraints on higher
order corrections and can be used as an order-consistency check of \PDF{} fits.
In this letter we use this connection to test the consistency of recent \PDF{} fits and propose its 
use in future \PDF{} fits. We also identify and discuss
the $b$-quark mass as a systematic uncertainty in the \PDF{}s that leads
to differences of more than five times the \PDF{} fit uncertainty.  It is a
systematic effect that has been broadly neglected so far but that has now become
dominant.

The novel consistency checks provided by this process are a result of the fact that, 
in \PDF{}s, bottom quarks are typically not included by fitting to data but instead
described perturbatively by splitting of the gluon \PDF{}.  In this ``five-flavor scheme''
the $b$-quark is an intrinsic part of 
the proton structure. Logarithms of the form $\alpha_s \log \left( (Q^2 + m_t^2)/m_b^2 \right)$, that appear
at each order of perturbation theory in $t$-channel single-top-quark production
(where $Q^2$ is a typical hard process scale and $m_b$ is the $b$-quark 
mass) are resummed through  evolution equations 
\cite{Olness:1987ep,Barnett:1987jw,Aivazis:1993pi,Stelzer:1997ns}.  For example,
the \LO{} expression for the $b$-quark PDF is,
\beq
b(x,\mu^2) = \frac{\alpha_s(\mu^2)}{2\pi}\ln \left(\frac{\mu^2}{m_b^2}\right) 
\int_x^1 \frac{dz}{z}
P_{bg}(z)g\left(\frac{x}{z},\mu^2\right) \,.
\label{eq:pdfevol}
\eeq
Successfully predicting 
cross-sections in the five-flavor scheme is therefore a strong test of 
our understanding of heavy quark \PDF{}s and the associated framework of improved perturbation 
theory.

A distinct property of this process is due to its structure:
the \LO{} Feynman diagram (shown in \cref{fig:tchan}) implies that the $W$-boson exchange 
factorizes
the process into  
two copies of \DIS{}. This factorization is robust under \QCD{} 
radiation for an on-shell 
top-quark, up to (small) interference corrections at \NNLO{}~\footnote{See e.g. 
	ref.~\cite{Bronnum-Hansen:2021pqc} for recent work in that direction.} so that higher-order effects 
	can be described 
by independent vertex corrections on the light-quark and heavy-quark lines. Consequently, this 
leads to an exact analytic
correspondence between the \DIS{} processes used to extract \PDF{}s from
data and the $t$-channel calculation.

When using \DIS{} data to extract \PDF{}s, the renormalization and factorization scales 
$\mu^2=Q^2$ are used, where $Q^2$ is the virtuality of the $W$ boson. Using any other scale choice 
does not 
necessarily reproduce the input data when \PDF{}s and matrix elements are combined
(though the difference diminishes with increasing perturbative
order). This fixes the scales for $t$-channel single-top to be
$\mu_l=Q^2$ on the light-quark line and $\mu_h=Q^2 + m_t^2$ on the heavy-quark line, which 
constitute the double-\DIS{} (\DDIS{}) scales.
It is exactly this particular choice of scales that effectively undoes \PDF{} fits and 
constrains total inclusive perturbative predictions to be the same at \LO{}, \NLO{} and \NNLO{}. 
The aforementioned interference and off-shell effects can alter this idealized picture. 
In addition, the inclusion of non-\DIS{} and neutral-current \DIS{} data in global fits allows
for more flexibility, effectively diluting this constraint.  Nevertheless 
the \DDIS{} constraint is essentially a strong correlation between the 
particular choice of scales used in the fitting procedure and the use of \PDF{}s in predictions. 
Such 
correlations have also been discussed in a broader context in 
refs.~\cite{Ball:2021icz,Harland-Lang:2018bxd}.

Taken together, precise predictions of this process in the five-flavor scheme therefore allow for a 
precision test of the heavy 
$b$-quark \PDF{} framework.
A previous study found that \LO{} and \NLO{} predictions for the Tevatron using 
various \PDF{} fits disagree by up to five times the \PDF{} fit uncertainty 
\cite{Sullivan:2017aiz}. It was suggested that such discrepancies
might just be an artifact of \LO{} fits that have not received the same
continuous attention as their higher-order counterparts.
To follow up on this hypothesis, we use our 
recent \NNLO{} calculation~\footnote{A first \NNLO{} calculation without top-quark decay has been 
	presented in 
	ref.~\cite{Brucherseifer:2013iv}. Later the decay has been added 
	\cite{Berger:2016oht,Berger:2017zof} and a discrepancy with the first calculation was found. 
	Subsequently, our calculation \cite{Campbell:2020fhf} resolved this discrepancy and added the 
	possibility to use different dynamic renormalization and factorization scales in both 
	production 
	pieces and the decay separately.}
that allows the use of 
\DDIS{} factorization and renormalization scales \cite{Campbell:2020fhf} to scrutinize 
the consistency of \LO{}, \NLO{} and \NNLO{} fits of different \PDF{} groups.
Small mistakes in either the \PDF{} calculation or input (e.g., 
through faulty evolution or poor fits) can reintroduce potentially large logarithms in
the calculation.
As a result,
formerly delicate cancellations -- that occur to enforce the equality of the inclusive
cross-sections between orders -- are broken, leading to large measurable deviations.

In the rest of this letter we use the framework described above as follows: We first check 
the order consistency of \PDF{} fits using \DDIS{} scales.
Focusing on \NLO{} and \NNLO{} here, we find that the commonly used modern \PDF{} fits are 
consistent between \NLO{} and \NNLO{}.
We then turn to the effect 
of the $b$-quark mass parameter and 
find that its uncertainty has been broadly neglected in \PDF{} fits. This leads to large 
systematic differences in $t$-channel single-top-quark predictions
between some commonly used \PDF{} fits. 
\NNLO{} predictions only agree within fit uncertainties once the 
systematic differences in $b$-quark masses are taken into account.
Since the momentum sum-rule connects heavy-quark \PDF{}s to other \PDF{}s, especially the gluon 
\PDF{}, it can also affect a range of other processes, albeit only at the subleading level.
We argue that on the path towards the proton structure at one-percent accuracy
\cite{Ball:2021leu} 
this systematic effect becomes relevant for \LHC{} phenomenology in general. In fact, 
for $t$-channel single-top-quark production it is now the dominant theoretical uncertainty.

\section{Order-consistency of modern PDFs}

Using \DDIS{} scales we expect that total inclusive cross-sections are equal at \LO{}, \NLO{} and 
\NNLO{}, up to small interference effects at \NNLO{} and up to non-\DIS{} data used in \PDF{} fits.
We perform our tests for Tevatron ($\sqrt{s}=\SI{1.96}{\GeV}$) $p\bar{p}$ collisions to have the 
highest sensitivity with respect to the \DDIS{} property. This is because for higher center of mass 
energies the 
differences between \LO{}, \NLO{} and \NNLO{} cross-sections shrink, regardless of the scales used, 
due to cancellations originating 
from the different parton density regimes. For 
example, in ref.~\cite{Sullivan:2017aiz} agreement to better than one percent between \LO{} and 
\NLO{} was found when using \DDIS{} scales and the {\abbrev CTEQ6} \PDF{} set. At the Tevatron the 
difference 
grows to $\sim10\%$ when using fixed $m_t$ scales instead, which allows for a stringent check. If
the 
\LHC{} was run at 
$\sqrt{s}=\SI{1.96}{\GeV}$ this difference would still be 
$9\%$, but an increase to $\sqrt{s}=\SI{8}{\TeV}$ or  $\sqrt{s}=\SI{13}{\TeV}$ reduces the 
sensitivity to $6\%$ for a top-quark and to $3\%$ for an anti-top-quark. To be maximally sensitive 
to the order consistency we therefore study Tevatron $p\bar{p}$ cross-sections at 
$\sqrt{s}=\SI{1.96}{\GeV}$. This also allows for a comparison with the measurements, in
principle, although we do not pursue that here. 

\begin{figure*}[htb]
	\includegraphics[width=\textwidth]{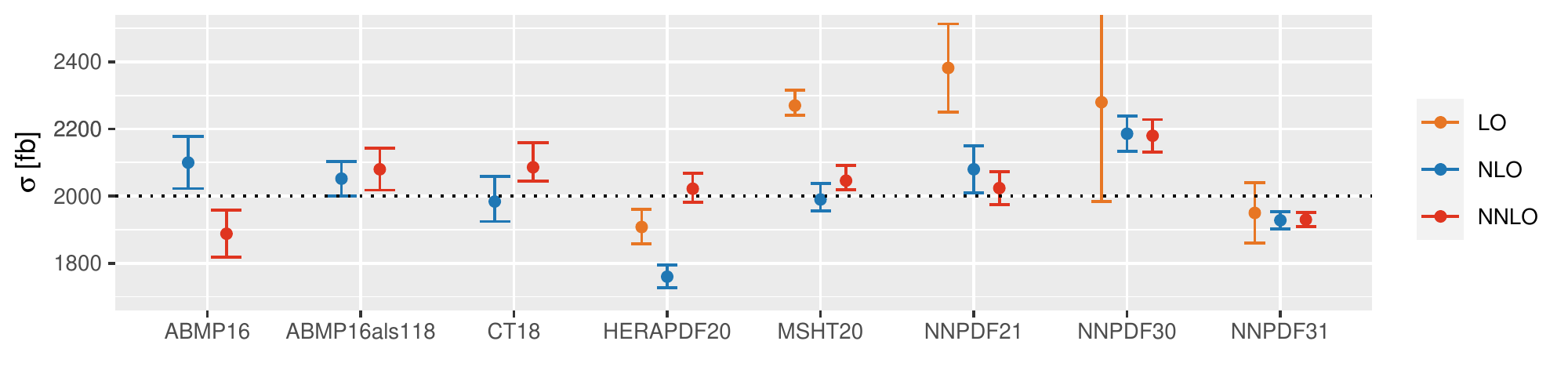}
	\caption{Inclusive Tevatron cross-sections for different \PDF{} sets at \LO{}, \NLO{} and 
		\NNLO{} with nominal 
		$m_b$ as used by each group. The solid error bars show  $1\sigma$ 
		\PDF{} fitting uncertainties.
	}
	\label{fig:consistency}
\end{figure*}

In \cref{fig:consistency} we show total inclusive $t$-channel cross-sections up to \NNLO{} using 
\DDIS{} scales with $1\sigma$ uncertainties as reported by \LHAPDF{} \cite{Buckley:2014ana} for 
various recent \PDF{} fits 
\cite{Alekhin:2018pai,Alekhin:2017kpj,Hou:2019efy,H1:2015ubc,Bailey:2020ooq,Ball:2011mu,NNPDF:2014otw,NNPDF:2017mvq}.
Note that only a few groups provide \LO{} fits with uncertainties and, with the exception
of {\abbrev NNPDF3.0} and {\abbrev NNPDF3.1}, there is no overlap between the uncertainties at \LO{}
and higher orders. We observe broad agreement 
between \NLO{} and \NNLO{} cross-sections.
For the {\abbrev ABMP} fits the \NLO{} \cite{Alekhin:2018pai} and \NNLO{} \cite{Alekhin:2017kpj} 
fits show a better consistency when using a fixed value of $\alpha_s(m_Z)=0.118$ (``{\abbrev 
ABMP16als118}''). The regular fit (``{\abbrev ABMP16}'') 
includes a fitted $\alpha_s(m_Z)\simeq0.119$ at \NLO{} and $\alpha_s(m_Z)\simeq0.115$ at \NNLO{}.
A fixed $\alpha_s(m_Z)$ must typically be treated as an additional systematic uncertainty and 
other groups commonly use the same value, at least at \NLO{} and \NNLO{}. 

On the other hand we observe significant large differences between the {\abbrev HERAPDF} 
\cite{H1:2015ubc} predictions, 
indicating a serious consistency issue. While one might expect other sets to deviate more 
between orders since they include non-\DIS{} data, this is not the case for {\abbrev HERAPDF}. The 
difference between \NLO{} and \NNLO{} is more than five times the \PDF{} fitting uncertainty.
Note that at the \LHC{} ($\sqrt{s}=\SI{13}{\TeV}$) the differences between \LO{}, \NLO{} and 
\NNLO{} completely disappear for {\abbrev HERAPDF}.

While the sensitivity at higher center of mass energies might be reduced, we suggest that it will 
still be useful to include \LHC{} single-top-quark data into fits with additional \DDIS{} 
constraints.

\section{$b$-quark masses in \PDF{} fits.}

We can also use \cref{fig:consistency} to compare the \NNLO{} predictions with each other.
A striking feature is the difference, greater than five times the fit uncertainty, between {\abbrev 
NNPDF3.0} \cite{NNPDF:2014otw} and the newer version 3.1~\cite{NNPDF:2017mvq}.
 The latter include new data and improved methodology, but a $5\sigma$ difference between 
 successive iterations is an unexpected feature.

The reason for this large difference lies in the widely different $b$-quark masses used in the 
fits. This was also found in ref.~\cite{Pagani:2020mov} where the authors studied $H$ and $Z$ 
associated single-top-quark 
production at \NLO{}. They find a cross-section difference that is several times larger than can be 
explained by  the
\PDF{} uncertainties of {\abbrev NNPDF3.0} and {\abbrev NNPDF3.1}, but state that the 
difference shrinks when using 
\NNLO{} \PDF{}s. In contrast, in our case the difference  observed at \NLO{} is maintained at 
\NNLO{}.

\begin{figure}[htb]
	\includegraphics[width=\columnwidth]{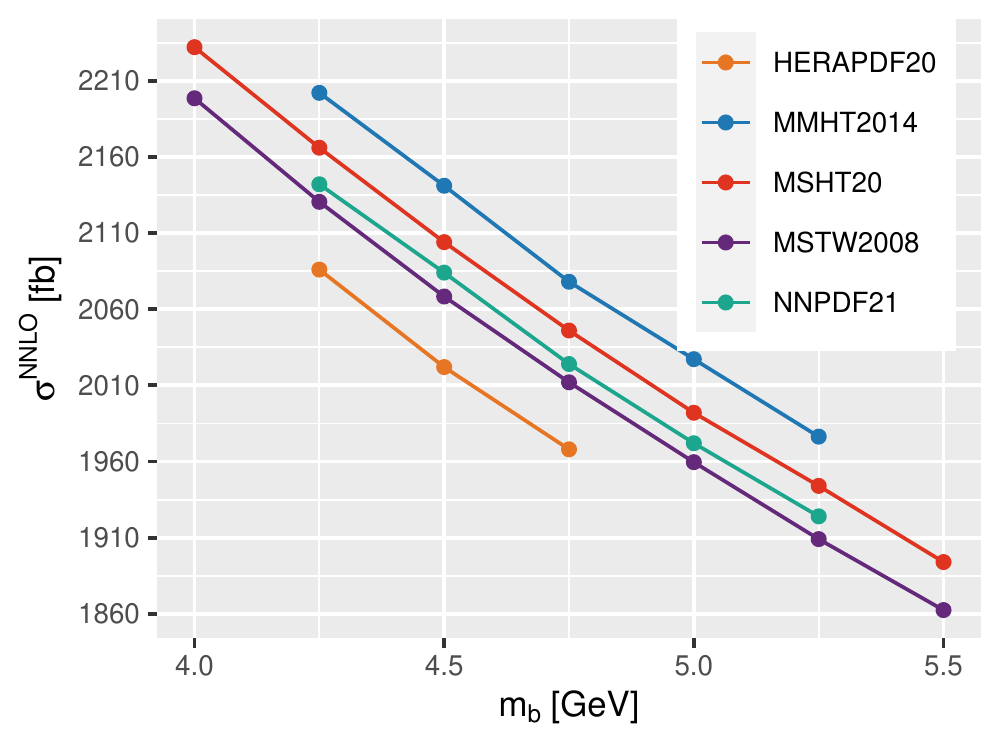}
	\caption{$b$-quark mass dependence of inclusive cross-section for different \NNLO{} \PDF{} 
		sets.}
	\label{fig:bmass}
\end{figure}

$t$-channel single-top-quark production and other heavy-flavor processes are largely dependent on 
the $b$-quark \PDF{}.
Their sensitivity to the $b$-quark mass is well-known in the 
\PDF{} community, with global analyses routinely emphasizing this
feature \cite{Martin:2010db,Ball:2011mu,Harland-Lang:2015qea,Cridge:2021qfd}. 
The overall finding is that the $b$-quark \PDF{} is strongly anticorrelated with the $b$-mass, 
i.e. when the mass is increased the $b$-quark \PDF{} is reduced.
We show this strong $b$-mass dependence in \cref{fig:bmass} for $t$-channel single-top-quark 
production. Varying the $b$-quark mass by \SI{0.2}{\GeV} results in a cross-section shift of about 
2\%. Another contributing systematic is the heavy-quark flavor scheme. It describes the 
matching of different calculations in the heavy-quark \PDF{} above and below the mass threshold 
\cite{Nocera:2019wyk}.

So while well-known in principle, we argue in the following that agreement on how to 
treat heavy-quark systematic effects are still 
incomplete and fragmented or outright missing at present.
An important contribution to the heavy-quark mass-induced uncertainties is the mass scheme used to encode it.
While the $\overline{\text{MS}}$ heavy-quark masses can be, and are, precisely determined 
\cite{PDG2020} the 
pole masses are unphysical. 
Furthermore, the matching series to schemes like the 
$\overline{\text{MS}}$ scheme diverges \cite{Beneke:2021lkq,Marquard:2015qpa}. Due to this issue,
typically a renormalon ambiguity of \SIrange{0.1}{0.2}{\GeV} is assumed 
\cite{Cridge:2021qfd,LHCHiggsCrossSectionWorkingGroup:2016ypw}. While the $\overline{\text{MS}}$ 
mass also has associated higher-order uncertainties, every \PDF{} fit 
working in the pole-mass scheme has an \emph{irreducible} mass uncertainty that 
translates to an additional systematic uncertainty of about 2\% in $t$-channel single-top-quark 
cross-sections. This is equal to or larger than the \PDF{} fit uncertainty of 
the most recent generation of \PDF{}s and larger than residual perturbative truncation 
uncertainties at \NNLO{}.

The theoretical framework for virtually all \PDF{} groups is based on the pole-mass scheme, and 
this is the limiting factor in the precision of current predictions of $t$-channel single-top-quark 
production. An exception to this is the {\abbrev ABMP} group --  the {\abbrev ABMP16} fit 
is consistently performed in the $\overline{\text{MS}}$ scheme 
\cite{Alekhin:2017kpj}. It furthermore directly includes the heavy-quark mass uncertainty by 
simultaneously fitting the heavy-quark masses.

All other \PDF{} groups fix the quark-mass in the fits. While that is a legitimate choice that can 
be made, unfortunately not all \PDF{} groups provide sets to 
vary the $b$-quark mass or give prescriptions to include heavy-quark mass uncertainties (see 
\cref{fig:bmass} for fits that allow for $b$-quark mass variation). For 
example the commonly used {\abbrev CTEQ} fits provide no variation sets and this systematic 
uncertainty is broadly neglected. The iterations of {\abbrev NNPDF} fits used recommended values 
from the literature with widely different values in different schemes. For example for 
{\abbrev NNPDF3.0}
it was argued \cite{NNPDF:2014otw} that, since scheme differences in their expressions are small, 
using a more precisely determined $\overline{\text{MS}}$ mass in place of a pole mass seemed appropriate.

\begin{figure*}[htb]
	\includegraphics[width=\textwidth]{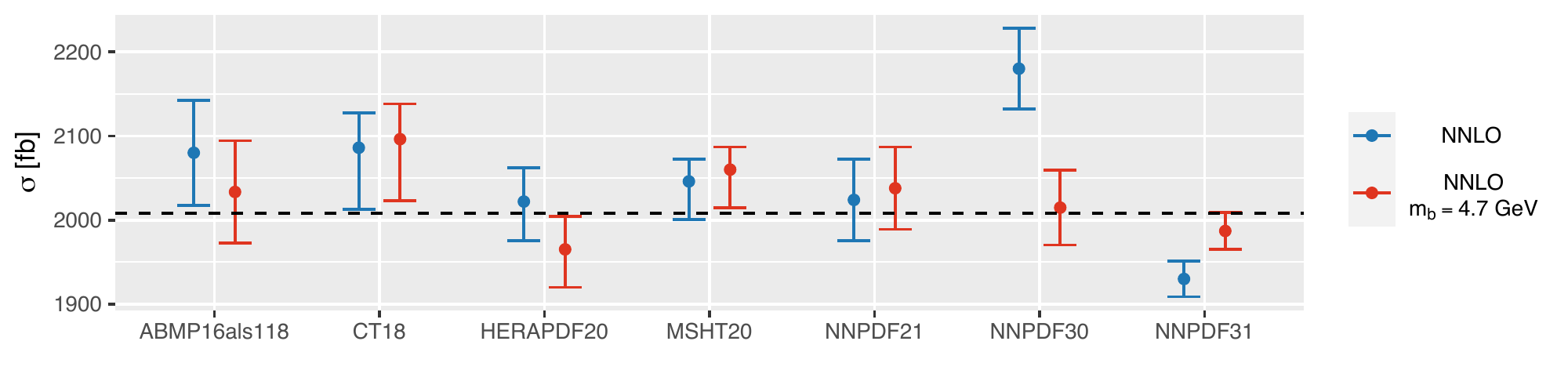}
	\caption{Inclusive Tevatron cross-sections for different \PDF{} sets at \NNLO{} with nominal 
		$m_b$ as used by each group, and rescaled to $m_b=4.7$~GeV. The solid error bars show $1\sigma$ 
		\PDF{} fitting uncertainties.
		The dashed horizontal line indicates a weighted average  of the 
		rescaled results (\SI{2007}{fb}). For the rescaling of the {\abbrev ABMP} results which use 
		the 
		$\overline{\text{MS}}$ mass in the fit we use its value converted to the pole scheme (see 
		text). }
	\label{fig:rescale}
\end{figure*}

Returning to the large difference found between {\abbrev NNPDF3.0} and {\abbrev NNPDF3.1}, it is 
straightforward
to check whether this difference can be explained by the choice of $m_b$. To do this, and also include 
all the other predictions on an equal footing, we have rescaled the 
\NNLO{} predictions by $\log(m_b)/\log(\SI{4.7}{\GeV})$  in \cref{fig:rescale}, where $m_b$ is the 
pole mass used in the individual \PDF{} fits.\footnote{
Note that {\abbrev ABMP} uses the 
$\overline{\text{MS}}$ mass in their 
fit, but to be able to compare with the other predictions we used their converted pole mass of 
$\SI{4.54}{\GeV}$ to rescale.}
At \LO{} this rescaling corresponds to an exact translation to a $b$-quark 
mass of \SI{4.7}{\GeV} since the process is purely $b$-quark initiated on the heavy-quark line,
c.f.\ \cref{eq:pdfevol}, and we 
expect this to be a good approximation at higher orders too \cite{Stelzer:1997ns}. The value of \SI{4.7}{\GeV} is chosen 
as an average of recently-used $b$-quark pole masses in \PDF{} fits between \SI{4.5}{\GeV} and 
\SI{4.9}{\GeV}. This rescaling indeed resolves the 
large $12\%$ discrepancy we found between using {\abbrev NNPDF3.0} and
{\abbrev NNPDF3.1}, where fixed (pole mass) values of 
$m_b=\SI{4.18}{\GeV}$ and $m_b=\SI{4.92}{\GeV}$  have been used, respectively.
For reference, the {\abbrev NNPDF3.0} fit uncertainty is $2\%$, while in {\abbrev NNPDF3.1} it 
decreases to $1.1\%$.
Moreover, after rescaling, all the predictions considered here are 
compatible within $\sim1\sigma$ fit uncertainties.

Given the large $b$-quark mass sensitivity, one would hope to use single-top-quark production 
to provide a measurement of the $b$-mass. Current uncertainties of $b$-quark masses 
obtained in \PDF{} fits are at the level of 2--3\% \cite{Alekhin:2017kpj} and matching this 
precision would 
require a large 
amount of statistics at the \LHC{}.  However, it would also require the elimination of systematic
uncertainties such as the  luminosity uncertainty. This would likely require the construction of a 
cross-section ratio, or normalized distribution, for which the $b$-mass dependence itself would
either be eliminated or strongly reduced. So, while $t$-channel single-top-quark production will 
help to constrain the $b$-quark mass in global \PDF{} fits, we do not anticipate its use as a 
measurement
channel on its own.

We conclude this section by noting that similar considerations also apply to the treatment of the 
charm-quark
mass in \PDF{} fits. Already in 
ref.~\cite{Alekhin:2017kpj,Accardi:2016ndt} it has been pointed out that fixing the charm-quark 
mass leads to the neglect of essential gluon \PDF{} correlations. In addition, then, using the
charm-quark 
mass in the pole-mass scheme with its large uncertainties leads to predictions with significant 
bias. 
For example 
the renormalon ambiguity in the charm-quark pole mass is about 10\%, while the 
$\overline{\text{MS}}$-mass can be obtained in \PDF{} fits to better than 2\% 
\cite{Alekhin:2017kpj}. For gluon 
fusion Higgs production at \NNLO{} this systematic uncertainty in the charm-quark mass translates 
into a cross-section uncertainty of $\pm 0.6\%$, the same order of magnitude as \PDF{} 
uncertainties of one 
to two percent reported by recent individual \PDF{} fits, see also \cite{Cridge:2021qfd}. Unless 
groups provide \PDF{} 
variation sets to estimate this systematic bias or include its effects directly in the fit, this 
represents a significant neglected uncertainty.

\section{Conclusions}

$t$-channel single-top-quark production strongly constrains order-by-order consistency of \PDF{} 
fits when using double-\DIS{} scales, especially when fits are dominated by \DIS{} data.  Focusing 
on differences between \NLO{} and \NNLO{}, we have 
tested several recent \PDF{} fits and find that most 
\PDF{}s are consistent across orders. One exception is the {\abbrev HERAPDF} fit, 
which shows a serious inconsistency between orders, indicating an issue. We propose 
to exploit the double-\DIS{} connection for future \PDF{} fits. 

On the other hand, comparing 
\NNLO{} cross-sections from different \PDF{} groups, we find that they are only consistent within standard 
fitting uncertainties once differences in $b$-quark masses are taken into account. 
With that we have identified and highlighted one of the most important systematic biases, the 
choice of mass entering the 
equations used to construct the heavy-quark \PDF{}s.
So far such effects have been broadly 
neglected, as evident by predictions based on \PDF{}s from the past ten years that lead to 
systematic 
differences of more than 10\%, but with \PDF{} fit uncertainties at the level of 1--2\%.
In fact for $t$-channel single-top-quark production we find that the 
$b$-quark mass uncertainty is currently the largest uncertainty overall. We further argued that to 
decrease this systematic uncertainty of the cross-section, the pole-mass scheme with its 
irreducible 
renormalon ambiguity of \SIrange{0.1}{0.2}{\GeV} will have to be abandoned in \PDF{} fits.

Heavy-quark mass effects are not just relevant for 
single-top quarks, but also for other measured processes like $Zb$ and $Zc$, which enter the LHC jet energy scales, and $Wbj$, which is a large background to multiple channels at the LHC. We encourage \PDF{} groups to update 
recommendations to include $b$-mass uncertainties and ideally avoid the pole mass ambiguity.
Our findings therefore strengthen the conclusions of ref.~\cite{Accardi:2016ndt}, 
where the authors argue that generally differences in \PDF{} sets are due to systematic effects.

\vspace{1em}
\begin{acknowledgments}

\paragraph{Acknowledgments.}

We would like to thank Frederick Olness for useful discussion
and Eleni Vryonidou for pointing out 
large differences due to the $b$-quark mass in $tZ$ and $tH$ 
production \cite{Degrande:2018fog}. 
This document was prepared using the resources of
the Fermi National Accelerator Laboratory (Fermilab), a
U.S. Department of Energy, Office of Science, HEP User
Facility. Fermilab is managed by Fermi Research Alliance, LLC (FRA),
acting under Contract No.\ DE-AC02-07CH11359. Tobias Neumann is supported by the United States 
Department of Energy under Grant Contract DE-SC0012704.
The numerical calculations reported in this paper were performed using the Wilson High-Performance 
Computing Facility at Fermilab.
\end{acknowledgments}
\bibliography{refs}

\end{document}